\documentstyle[floats,twocolumn,prl,aps]{revtex}
\begin{document}
\draft
\def\ref{\par\noindent\hangindent=3mm\hangafter=1}
\narrowtext
{
\title{
The gap symmetry is marginal in high-$T_c$ superconductivity:
Isotropic s-wave pairing
leads to either s or d-wave gap
depending on the Coulomb pseudopotential}
}
\author{G. Varelogiannis}
\address{
Dipartimento di Fisica,
Universit\`a di Roma ``La Sapienza'', Piazzale Aldo Moro 2, I-00185 Roma,
Italy}
\author{\parbox{397pt}{\vglue 0.3cm \small
We point out that in two-dimensional systems
like high-$T_c$ superconductors,
an isotropic s-wave electron-phonon interaction in the momentum decoupling
regime leads to either s or d-wave superconductivity
depending on the conventional values of the Coulomb pseudopotential.
It is therefore possible that $YBa_2Cu_3O_7$ is
an anisotropic
d-wave and $Bi_2Sr_2CaCu_2O_8$ an anisotropic s-wave
superconductor both having the same pairing interaction and similar
$T_c$'s. A transition from s-wave
to d-wave superconductivity may also be possible by adjusting
the doping.
}}
\maketitle
\par
The symmetry of the order parameter is in the center of the debate
on the origin of high-$T_c$ superconductivity \cite{Dynes}.
It is generally believed that the symmetry of the gap
should also indicate the symmetry of the pairing mechanism.
Studying this symmetry should allow one to determine
whether cuprates are conventional s-wave superconductors
with an interaction that could be mediated by phonons, or
whether they are unconventional
d-wave superconductors in which case the
pairing interaction should be due to spin fluctuations \cite{spin}
or to some other exotic mechanism. In this spirit one expects that
$Bi_2Sr_2CaCu_2O_8$ and $YBa_2Cu_3O_7$ should both have the same symmetry,
and their high critical temperatures should be intimately related
to the symmetry of the gap.

Qualitative aspects
of anisotropy in $Bi_2Sr_2CaCu_2O_8$,
like
the correlation between
the gap, the electronic density of states (DOS)
and the visibility of the dip above the gap anisotropies \cite{Shen},
the strong enhancement of the anisotropy with temperature \cite{Onellion},
the asymmetry of the SIN tunnel spectra where the dip is seen only at
negative sample bias \cite{Renner} and also the possible presence of
a secondary gap maximum or ``hump'' in the $\Gamma-X$
direction \cite{Capunzano}, have been shown to
indicate that the pairing interaction is isotropic s-wave in the
momentum decoupling regime \cite{Previous}.
Momentum decoupling (MD) arises when the characteristic momenta exchanged
during the pairing interaction are small compared to the characteristic
momenta of the electronic density of states (DOS) variations over the
Brillouin zone. In that case the Eliashberg equations in the momentum space
tend to decouple and have different couplings
in different regions of the Fermi surface.
These couplings are proportional to the local DOS and therefore anisotropies
of superconductivity reflect the DOS anisotropies.
Our analysis of the $Bi_2Sr_2CaCu_2O_8$ phenomenology in this context
points to an anisotropic
s-wave gap \cite{Previous}.
However, various phase sensitive \cite{phase1} and node sensitive
\cite{nodes} experiments on $YBa_2Cu_3O_7$ report evidence of
a sign reversal of the order parameter
supporting d-waves \cite{Mazin}, although experimental
contradictions
persist \cite{Chaudhari,reviewSD}.


We will show in this letter
that, contrary to the general belief,
the symmetry of the order parameter does not represents a crucial
criterion for the symmetry of the pairing interaction, and
is an energetically marginal parameter.
We will see in particular
that an isotropic s-wave
interaction
(that could be mediated
by phonons) in the momentum decoupling regime
can lead to {\it either} s-wave or d-wave superconductivity depending on
parameters that are {\it marginal} for the pairing mechanism.
These parameters are the
magnitude of the Coulomb pseudopotential $\mu^*$ and the
relative importance of the characteristic momenta of the variations
of the Coulomb pseudopotential $\mu^*$
compared to the characteristic momenta
exchanged during the pairing interaction.
It is for example possible
that $Bi_2Sr_2CaCu_2O_8$ is an anisotropic s-wave
and $YBa_2Cu_3O_7$ an {\it anisotropic} d-wave superconductor
both having {\it the same}
attractive isotropic s-wave interaction in the momentum
decoupling regime. In addition, it may be possible to
switch from s-wave to d-wave superconductivity and vice versa
by doping since $\mu^*$ depends sensitively on the doping.
Of course
it will also be clear that a d-wave gap symmetry {\it does not} imply
a spin fluctuation pairing mechanism \cite{spin}.

Energetic irrelevance
of the symmetry of the order parameter has also been reported
in a model proposed to describe the
``spin gap'' in underdoped cuprates \cite{millis}.
This property is also seen in another model
in which the Fermi surface is divided
in three independent pieces and the interaction has
a low energy cut-off \cite{Santi}.
The possibility that phonons may help d-wave superconductivity which is
primarily due to spin fluctuations has been reported in Ref.
\cite{Lichtenstein}, while more recently Nazarenko and Dagotto \cite{Nazarenko}
proposed a specific Holstein model with nearest neighbor attraction
which leads to d-wave superconductivity the spin fluctuations playing a
secondary role.

The Coulomb pseudopotential $\mu^*$ represents the
effective repulsion between the paired electrons \cite{Allen} and
is present in all conventional superconductors,
acting essentially as an effective negative coupling \cite{maxinf} and
reducing slightly the isotope effect exponent.
In a two dimensional system in the MD regime $\mu^*$
becomes a crucial parameter having unexpected implications
on the gap symmetry.
We performed simulations on a
BCS model, in which case the gap is given by (we use the notations
of \cite{Previous})
$$
\Delta(\vec{k})=\sum_{\vec{p},|\xi_p|<\Omega_D}
{-\bar{V}(\vec{k}-\vec{p})
\Delta(\vec{p})\over 2\sqrt{\xi^2_{\vec{p}}+\Delta^2(\vec{p})}}
\tanh\biggl(\sqrt{{\xi^2_{\vec{p}}+\Delta^2(\vec{p})\over 2T}}\biggr)
$$
Where $\bar{V}(\vec{q})=V^A(\vec{q})+\mu^*(\vec{q})$ is
the sum of the attractive interaction, due for example
to electron-phonon coupling $V^A$, and the effective Coulomb repulsion $\mu^*$.
We solve this equation iteratively on a grid $120\times 120$
in the zero temperature regime.

We consider an isotropic s-wave electron-phonon
coupling having at small momenta a Lorentzian
behavior as a function of the norm of the exchanged momentum
$
V^A(\vec{q})=-g^2\bigl(1+|\vec{q}|^2/Q_c^2(e-ph)\bigr)^{-1}
$.
In this spectrum the electron-phonon scattering is dominated
by the processes which transfer a momentum smaller than $Q_c(e-ph)$.
In our analysis $Q_c(e-ph)$ is the relevant parameter and the
particular shape of the interaction is irrelevant.
As for the repulsive interaction,
we first consider the simple case of a momentum independent Coulomb
pseudopotential $\mu^*(\vec{q})=\mu^*_0$
(hard-core-like yet finite repulsion).
Band structure effects are completely marginal for our discussion here,
and for clarity we show results corresponding to
the
simple nearest neighbor tight binding dispersion
at half-filling $\xi_{\vec{k}}=-2t[\cos(k_x)+\cos(k_y)]$
(the lattice spacing is taken equal to unity).
All our discussion is unaltered by the inclusion of next nearest
neighbor or other terms in the dispersion relation \cite{later}.

We show in figure 1 some of the calculated momentum dependent
gap functions on the Fermi
surface, for different values of $\mu^*_0$
in both the d-wave and s-wave channel.
When $k$ varies from $0$ to $\pi$
we cover a quadrant of the
Fermi surface
Here we take
$Q_c(e-ph)\approx\pi/12$ which places us
in the momentum decoupling regime
and is relevant for
$Bi_2Sr_2CaCu_2O_8$ \cite{Previous}. The momentum dependence of the gap in the
s-wave channel at zero repulsion (upper full line)
is precisely due to momentum decoupling.

When we introduce repulsion in the s-wave channel, the gap is
reduced by a constant amount in all directions (triple-dot-dashed line)
resulting therefore in an effective
enhancement of the anisotropy. At a critical
value of the repulsion which in the case considered in Fig. 1 is
on the order $\mu^*/g^2\approx 0.21$ (dot-dashed line), the gap becomes
almost zero in the $(\pi,\pi)$ direction, and we have a
{\it discontinuous} transition to a new gap symmetry structure
with two nodes in the quadrant shown in Fig. 1
(dotted and dashed lines).
In this new state the gap becomes {\it independent} on the magnitude
of $\mu^*$ and the areas of the Fermi surface at which the gap is
positive are equal to the areas in which the gap is negative.
We will see in the following that this new state is not the physical
state occupied by the system because at these $\mu^*$ the d-wave
solution is energetically more favorable.

The d-wave solution is also characterized by
independence of the gap on the magnitude of $\mu^*$.
The origin of the repulsion independence of the gap lies
on the momentum independence of $\mu^*$,
on the equality of the Fermi surface areas with positive and
negative gap, and on the two dimensional character of the
system considered here.
We can see this analytically by considering a circular Fermi surface.
If the gap on the Fermi surface has the form
$\Delta_{\vec{k}}=\Delta\cos(n\phi)$ where $n$ is the number of
nodes and $\phi$ is the polar angle in the usual polar coordinates,
then the momentum independent repulsive contribution to the gap function
becomes proportional
to the integral \cite{Peter}
$$
I=\int_0^{2\pi}d\phi {\Delta\cos(n\phi)\over
\sqrt{\xi^2_{|\vec{k}|}+\Delta^2\cos^2(n\phi)}}=0
$$
which is {\it identically} zero. Therefore a
gap changing sign periodically
on a two dimensional system eliminates the effect of any
{\it momentum independent} repulsion.

We obtain in this way a qualitative understanding of the
discontinuous nature of the transition in the s-wave channel in Fig. 1
(from dot-dashed live at $\mu^*/g^2=0.21$ to dashed line at $\mu^*/g^2=0.22$).
We also remark on the following points that might generally
apply to any two-dimensional system:
First, it is impossible to get a zero gap somewhere
on the Fermi surface without a variation of sign of the order
parameter. Secondly, the system can
only have nodes with a finite slope
and therefore if there are nodes a linear $T$-dependence
of the penetration depth at low temperatures is plausible \cite{nodes}.
Finally, when the gap changes sign, the area of the Fermi surface
in which it is positive is equal to the area in
which it is negative.

In realistic situations
$\mu^*$
is momentum dependent (the repulsion is not hard-core-like)
To introduce a
smooth momentum cut-off for the variations of $\mu^*$
we consider a
structure analogous to that of the pairing interaction
written at small $q$ as
$
\mu^*(\vec{q})=\mu^*_0\bigl(1+|\vec{q}|^2/Q_c^2(Cb)\bigr)^{-1}
$, in which case $Q_c(Cb)$ represents the characteristic range
of the exchanged momenta (smooth cut-off) in the repulsive interaction.
We considered various different momentum structures of $\mu^*$
and we checked that the
following discussion remains valid with
only slight quantitative modifications \cite{later}.
The important parameter is the momentum cut-off in the repulsive
interaction $Q_c(Cb)$ compared to that
in the attractive pairing interaction $Q_c(e-ph)$.
We show in Fig. 2 the s-wave and d-wave gap solutions for
$Q_c(Cb)=\pi/4$ and $Q_c(e-ph)=\pi/12$. Now the d-wave solution is not
repulsion independent but it still appears less sensitive
to the repulsion than the s-wave solution.
It is clear from Fig. 2 (and also Fig. 1) that at some critical repulsion
the absolute d-wave gap will become larger in average than
the s-wave gap, and this will become the
energetically favorable state.

To find out which of the
s-wave and d-wave solutions is the physical state of the
system, one has to calculate the free energy gain due to the
superconducting transition \cite{Bardeen},
the physical solution being that with the higher absolute
free energy gain (the lower free energy).
Of course the solution with the higher free energy gain
is that for which the integral of the absolute value of the gap
on the Fermi surface is higher
and this is also the solution with the higher $T_c$.
In figure 3 we show the evolution of the free energy gain
(absolute value of the free energy due to the superconducting transition
in arbitrary units)
as a function of $\mu^*_0/g^2$ in the
case $Q_c(Cb)=\pi/4$ and $Q_c(e-ph)=\pi/12$. The s-wave
solution is favorable when $\mu^*_0/g^2$ is small but when
this ratio takes values larger than a critical value of the
order of $\mu^*_0/g^2\approx 0.15$
then the d-wave solution
becomes more favorable. If we
were able to control $\mu^*$
for example by doping then we could
induce transitions
from s-wave
to d-wave superconductivity and vice versa.

Notice that such s-d transitions occur in our system at
conventional values of the Coulomb pseudopotential already seen in
low-$T_c$ superconductors \cite{Allen}.
The critical $\mu^*_0$ is very dependent on the ratio $Q_c(e-ph)/Q_c(Cb)$,
We studied the evolution of the
free energy gain in the s and d-wave channels
as a function of $\mu^*_0/g^2$
for different values of $Q_c(e-ph)/Q_c(Cb)$.
This allowed us to construct the phase diagram shown in Figure 4.
We also made an analogous study in the case of different momentum structures
of $\mu^*$ having an effective
momentum cut-off and the resulting phase diagrams are similar to that
of Fig. 4 \cite{later}.
When for a given $Q_c(e-ph)/Q_c(Cb)$
the ratio $\mu^*/g^2$ is larger than the critical value
showed in Fig. 4, the gap is d-wave.
It is clear that the smaller the ratio $Q_c(e-ph)/Q_c(Cb)$ the smaller is
the value of the critical repulsion while for
$Q_c(e-ph)$ of the same order as $Q_c(Cb)$ the d-wave solution is impossible
and this is probably the case in low-$T_c$ metallic superconductors.
Since in the MD regime $Q_c(e-ph)\ll k_F$, we can also have
$Q_c(e-ph)\ll Q_c(Cb)$, in which case s and d-wave superconductivity are
energetically close and both states are physically acceptable
depending on the precise value and structure of $\mu^*$.
We remark that if in our scheme
we want to have both d-wave and high-$T_c$, then it is better to
be well in the MD regime and have a small critical $\mu^*$,
since large
$\mu^*$ reduces $T_c$.
Notice that the evolution of $T_c$ follows essentially
that of the free energy gain and
from figures 2 and 3 one can conclude
that the negative effect of $\mu^*$ on $T_c$ is smaller
in the case of d-waves than in the case of s-waves \cite{later}.

It is important to notice that the d-wave state we obtain is anisotropic
and its anisotropies are driven by the DOS anisotropies in the same way as
for the s-wave state \cite{Previous}.
This is clear in Figures 1 and 2 where the d-wave solution
away from the $(\pi/2,\pi/2)$ direction has exactly the form of the s-wave
solution.
When we are in the MD regime we always have
different couplings
in different regions of the Fermi surface whatever the symmetry of the gap
and a large part of our
discussion of ARPES and tunneling in $Bi_2Sr_2CaC_2O_8$ \cite{Previous}
remains valid even for the d-wave state. In particular all our
discussion on the origin of the dip structure \cite{RCdip},
the correlation of gap, dip and DOS anisotropies, and the asymmetry
of SIN tunnel spectra \cite{Previous} is also true in the MD d-wave.
We remark that in the case of a d-wave state obtained by spin
fluctuations it was impossible to reproduce the dip above the gap
by strong coupling calculations \cite{Scalapino}.

We are grateful to Prof. L. Pietronero for stimulating discussions
and encouragement.



{\bf Figure 1:} The gap as a function of $k_x$ (or $k_y$)
for a momentum independent repulsion $\mu^*$ and
$Q_c(e-ph)=\pi/12$. We consider
the s-wave channel and $\mu^*=0$ (upper full line), $\mu^*/g^2=0.05$
(triple-dot-dashed line), $\mu^*/g^2=0.21$ (dot-dashed line),
$\mu^*/g^2=0.22$ (dotted line), $\mu^*/g^2=0.30$ (dashed line) and the
d-wave channel for the same $\mu^*/g^2$ ratios (lower full line).
The d-wave solution is independent of $\mu^*$.

{\bf Figure 2:} The s-wave gap (full lines) and the d-wave gap
(dashed lines) as a function of $k_x$ (or $k_y$)
on a quadrant of the Fermi surface,
for three different
values of the repulsion $\mu^*_0/g^2=0.05$, $0.15$ and $0.25$,
in the case $Q_c(e-ph)=\pi/12$ and $Q_c(Cb)=\pi/4$.
The smaller
the ratio $\mu^*_0/g^2$ the larger is the absolute value of the gap in
both the s and d-wave solutions.

{\bf Figure 3:} The absolute value of the free energy gain
due to the superconducting transition as a function of the
ratio $\mu^*_0/g^2$ for $Q_c(e-ph)=\pi/12$ and $Q_c(Cb)=\pi/4$.
The dotted line (triangles) corresponds to the d-wave solution and the
full line (circles) to the s-wave solution. The physical solution is that
with the higher absolute free energy gain (lower free energy).

{\bf Figure 4:} Phase diagram. The critical repulsion $\mu^*_0/g^2$
for the transition from s-wave to d-wave superconductivity
as a function of the ratio $Q_c(e-ph)/Q_c(Cb)$. The upper region of
the graph corresponds to the d-wave state.


\begin{thebibliography}{999}

\bibitem{Dynes} See for example: B. Goss Levi, Phys. Today {\bf 46}(5),
17 (1993)

\bibitem{spin} N.E. Bickers, D.J. Scalapino and S.R. White,
Phys. Rev. Lett. {\bf 62}, 961 (1989);
P. Monthoux, A. Balatsky and D. Pines, Phys. Rev. Lett. {\bf 67},
3348 (1991)



\bibitem{Shen} Z.-X. Shen and D.S. Dessau, Physics Reports {\bf 253},
p. 1-162 (1995)

\bibitem{Onellion} J. Ma, C. Quitmann, R.J. Kelley, H. Berger,
G. Margaritondo, and M. Onellion, Science {\bf 267}, 862 (1995)

\bibitem{Renner} Ch. Renner and \O. Fischer, Phys. Rev. B {\bf 51},
9208, (1995)

\bibitem{Capunzano} H. Ding et al., Phys. Rev. Lett. {\bf 74},
2784 (1995)

\bibitem{Previous} G. Varelogiannis, A. Perali, E. Cappelluti and
L. Pietronero, preprint cond-mat/9507052

\bibitem{phase1} D.A. Wollman, D.J. Van Harlingen, W.C. Lee,
D.M. Ginzberg and A.J. Leggett, Phys. Rev. Lett. {\bf 71},
2134 (1993);
D.A. Brawner and H.R. Ott, Phys. Rev. B {\bf 50},
6530 (1994);
C.C. Tsuei, J.R. Kirtley, C.C. Chi, L.S. Yu-jahnes,
A. Gupta, T. Shaw, J.Z. Sun, and M.B. Ketchen, Phys. Rev. Lett. {\bf 73},
593 (1994)

\bibitem{nodes} W.N. Hardy, D.A. Bonn, D.C. Morgan, R. Liang and K. Zhang,
Phys. Rev. Lett. {\bf 70}, 3999 (1990);
L.A. de Vaulchier, J.P. Vieren, Y. Guldner, N. Bontemps,
R. Combescot, Y. Lemaitre and J.C. Mage, preprint cond-mat/9504062

\bibitem{Mazin} Some models tend to describe the phase sensitive
experiments in $YBa_2Cu_3O_7$ without d-waves:
I.I. Mazin, A.A. Golubov and A.D. Zaikin,
Phys. Rev. Lett. {\bf 75}, 2574 (1995);
D.Z. Liu, K. Levin, and J. Maly, Phys. Rev. B {\bf 51},
8680 (1995)

\bibitem{Chaudhari} P. Chaudhari and S.-Y. Lin, Phys. Rev. Lett.
{\bf 72}, 1084 (1994);
A.G. Sun, D.A. Gajewski, M.B. Maple and R.C. Dynes,
Phys. Rev. Lett. {\bf 72}, 2267 (1994)

\bibitem{reviewSD} For reviews see: R.C. Dynes, Sol. State Commun. {\bf 92},
53 (1994); J.R. Schrieffer, Sol. State Commun. {\bf 92}, 129 (1994)

\bibitem{millis} B.L. Altshuler, L.B. Ioffe and A.J. Millis,
To be published in Phys. Rev. B. We thank Andy Millis for bringing this
reference to our attention and for a very stimulating discussion

\bibitem{Santi} G. Santi, T. Jarlborg, M.Peter and M. Weger,
J. of Supercond., {\bf 8}, 405 (1995)

\bibitem{Lichtenstein} A.I. Lichtenstein and M.L. Kulic, Physica C
{\bf 245}, 186 (1995)

\bibitem{Nazarenko} A. Nazarenko and E. Dagotto, preprint cond-mat/9510030

\bibitem{Peter} A model with a cylindrical Fermi surface is studied in:
M. Peter and G. Varelogiannis, preprint

\bibitem{Allen} P.B. Allen and B. Mitrovic, in Solid State Physics {\bf 37},
(1981); J.P. Carbotte, Rev. Mod. Phys. {\bf 62}, 1027 (1990)

\bibitem{maxinf} G. Varelogiannis, Physica C {\bf 249}, 87 (1995)

\bibitem{later} A detailed analysis will be published elsewhere

\bibitem{Bardeen} J. Bardeen and M. Stephen, Phys. Rev. {\bf 136},
A1485 (1964)

\bibitem{RCdip} G. Varelogiannis, Phys. Rev. B {\bf 51}, 1381 (1995)

\bibitem{Scalapino} D.J. Scalapino, Private Communication,
see also P. Monthoux and D.J. Scalapino, Physica B {\bf 199-200},
294 (1994)



\end{thebibliography}
\end{document}